\documentstyle[prl,aps,preprint]{revtex}
\begin{document}

\title{ Many Body Correlation Corrections to Superconducting
Pairing in Two Dimensions}

\author {M.Yu. Kuchiev\cite{Ioffe} and O. P. Sushkov\cite{Budker}}
\address { School of Physics, The University of New South Wales
 Sydney 2052, Australia}

\date{March 31, 1995}
\maketitle

\begin{abstract}
We demonstrate that in the strong coupling limit (the superconducting
gap $\Delta$ is as large as the chemical potential  $\mu$),
which is relevant to the
high-$T_c$ superconductivity, the correlation corrections to the
gap and critical temperature are about 10\% of the corresponding mean
field approximation values. For the weak coupling
($\Delta \ll \mu$) the correlation corrections are very large: of
the order of 100\% of the corresponding mean field values.
\end{abstract}

\vspace{0.5cm}
\hspace{3.cm}PACS: 74.20.-z, 74.20.Fg, 74.20.Mn
\vspace{1.cm}

\section{Introduction}
In recent papers \cite{Fla4,Bel5} we demonstrated
the very strong  d-wave pairing between dressed quasiholes in the
$t-J$ model induced by   spin-wave exchange.
The pairing gives the critical
temperature in a reasonable agreement with experimental data for Cooper
Oxide Superconductors. For calculations we used BCS-like mean field
approximation for dressed quasiholes.  A similar approach to the
pairing of dressed quasiholes has been used by Dagotto, Nazarenko, and
Moreo \cite{Naz}. The important difference is that in \cite{Fla4,Bel5}
the hole-hole interaction was derived from the parameters of the
t-J model, while in \cite{Naz} it was introduced  {\it ad hoc}
with magnitude adjusted to fit experimental data.
The typical
value of the gap $\Delta$ obtained in the papers \cite{Fla4,Bel5,Naz}
is of the order of the chemical potential $\mu$. This strong pairing
inspires  a natural question : how strongly the hole-hole
correlations influence upon  the mean field result?
The purpose of  the present work is to investigate
this problem.
We consider the simplified model instead of the sophisticated t-J one.
The model under consideration describes  the two-dimensional fermions with
quadratic dispersion and contact attraction. This model
inherits the main, relevant to the problem, property of the t-J model:
it permits to consider strong pairing.
Therefore one can suppose the obtained result to be rather general.
We study the dependence of correlations on
the intensity of attractive interaction.   Our conclusion
is that for strong coupling $\Delta \ge \mu$ the correlation
corrections are not large: about 10\% of the corresponding mean field value.
This is a very surprising result since one could expect that  strong
interaction causing the pairing might make the correlation correction to
be very large. Our conclusion is in agreement with the one which was
recently obtained  in finite-cluster numerical study of $t-J$ model by
Ohta, Shimozato, Eder, and Maekawa \cite{Ohta}.

 The other surprising result of the present work is that in the weak
coupling limit $\Delta \ll \mu$ the correlation corrections are very large:
the renormalizations of  mean field values are about 100\%. This is due
to the specific ultraviolet behaviour in the two dimensional theory with
attraction.

 We are mainly interested in the regime $\Delta \ge \mu$ when the system is
close to  a
smooth crossover from a state with large, overlapping Cooper pairs to
a Bose condensate of composite bosons. In the mean field approximation
this crossover has been studied in three dimensions by Legget\cite{Legget}
and by Nozieres and Schmitt-Rink\cite{Noz}.
A similar problem for two dimensions
reveals interesting features considered by
Randeria, Duan, and Shieh\cite{Ran}. In the present work we investigate the
correlation corrections to the results presented in Ref.\cite{Ran}

\section{Mean Field Approximation}
Consider the Hamiltonian of fermions with spin $1/2$ and contact attractive
spin independent interaction
\begin{equation}
\label{H}
H=\sum_{{\bf k},\sigma}{{{\bf k}^2}\over{2m}}
a_{{\bf k},\sigma}^{\dag} a_{{\bf k},\sigma}
-{g\over V}
\sum_{\bf k_1,k_2,k_3} a_{{\bf k_1+k_2-k_3},\uparrow}^{\dag}
a_{{\bf k_3},\downarrow}^{\dag} a_{{\bf k_2},\downarrow}
a_{{\bf k_1},\uparrow},
\end{equation}
where summation over ${\bf k}$ is restricted in two-dimensional
plane, $V$ is the area of the plane,
$\sigma= \pm 1/2 = \uparrow, \downarrow$
is a  projection of the usual 3-dimensional spin 1/2. Consider first the
Schroedinger equation for a two-particle bound state with zero total momentum
\begin{equation}
\label{Sch}
\chi_{\bf k}=-{g\over V}\sum_{\bf p}{{\chi_{\bf p}}\over{E_a-{\bf p}^2/m}}.
\end{equation}
The sum here is logarithmically ultraviolet divergent, and therefore
one has to introduce the ultraviolet cutoff
$E_{\Lambda}=\Lambda^2/2m$. For the $t-J$ model the parameter $\Lambda$ is
of the order of inverse lattice spacing. Solution of Eq.(\ref{Sch})
is straightforward
\begin{equation}
\label{BSt}
\chi_{\bf p}=const, \hspace{0.5cm}
E_a=-2E_{\Lambda}e^{-4\pi /mg}.
\end{equation}
Now let us consider a many body problem with fixed number density of
particles $\delta$. The Fermi energy is: $E_F=\pi \delta/m$.
The BCS equation for the pairing at fixed chemical potential $\mu$
is of the form
\begin{equation}
\label{BCS}
\Delta_{\bf k}={g\over V}\sum_{\bf p} {{\Delta_{\bf p}}\over
{2 \epsilon_{\bf p}}}
\tanh {{\epsilon_{\bf p}}\over {2T}},
\end{equation}
where $\epsilon_{\bf p}= \sqrt{\eta_{\bf p}^2+\Delta_{\bf p}^2}$,
and $\eta_{\bf p}={\bf p}^2/2m - \mu$. Similarly to the Schroedinger
equation we have to introduce the ultraviolet cutoff $\Lambda$.
Solution of the BCS equation for zero temperature is straightforward.
Assuming that $E_F \ll E_{\Lambda}$ one gets
\begin{eqnarray}
\label{RDS}
\Delta&=&\sqrt{2E_F|E_a|}=2 \sqrt{E_FE_{\Lambda}} e^{-2\pi /mg},\\
\mu&=&E_F-|E_a|/2=E_F\biggl(1-{1\over
4}{{\Delta^2}\over{E_F^2}}\biggr),\nonumber
\end{eqnarray}
where $E_a$ is the binding energy (\ref{BSt}) of the two particle bound
state. This solution obtained by Randeria, Duan, and Shieh \cite{Ran} gives
a smooth crossover from the BCS limit ($\mu \approx E_F$)
to the Bose condensate of composite bosons ($\mu < 0$). In the
present work we concentrate on the case of positive $\mu$ because,
in our opinion, it is relevant to the realistic high-$T_c$
superconductors.

 The critical temperature $T_c$ can be easily found from Eq.(\ref{BCS})
if we define it as a point where the gap vanishes. In the weak
coupling limit ($T_c \ll E_F$) we have the usual BCS relation:
$\Delta(T=0)/T_c=1.76$. Numerical solution of Eq.(\ref{BCS})
shows that even in the strong coupling limit this ratio remains
very close to the BCS value. For example at $\Delta(T=0)/\mu(T=0)=3$
which is equivalent to $\Delta(T=0)/E_F \approx 1.5$
one finds $\Delta(T=0)/T_c=1.68$. We assume that the density of
particles is fixed. One has to remember that under this condition
the chemical potential is a function of temperature:
$\mu(T=0) \ne \mu(T=T_c)$. For strong coupling the temperature
dependence of the chemical potential is not negligible.

 Equation (\ref{BCS}) as well as  solution (\ref{RDS})
describe the mean field approximation. Now let us consider
correlation corrections.
We use the conventional Gorkov-Nambu technique, see e.g. book \cite{Land}.
Consider first the case of zero temperature.

\section{Correlation Correction to the Gap at Zero Temperature}

The normal $G(1,2)=-i\langle T[\psi(1) \psi^{\dag}(2)]\rangle$ and
the anomalous
$F^{\dag}(1,2)=-i\langle T[\psi^{\dag}(1) \psi^{\dag}(2)]\rangle$
Green functions obey the usual Dyson equations \cite{Land}
\begin{eqnarray}
\label{D}
\hat G(p)&=&[1+\hat G(p) \hat \Sigma_{11}(p) +
\hat F^{\dag}(p) \hat \Sigma_{20}(p)] \hat G_0(p) \\
\hat F^{\dag}(p)&=&[\hat F^{\dag}(p) \hat \Sigma_{11}(-p)+
\hat G^(p) \hat \Sigma_{02}(p)] \hat G_0(-p),\nonumber
\end{eqnarray}
where $p=(\epsilon, {\bf p})$,
$G_0(p)=[\epsilon-\eta_{\bf p}+i0\cdot sign(\eta_{\bf p})]^{-1}$.
In the first order of
perturbation theory the normal self-energy operator $\hat \Sigma_{11}^{(1)}$
is given by the diagrams presented in Fig.1. With the interaction
(\ref{H}) the self-energy $\hat \Sigma_{11}^{(1)}$ is momentum independent,
and therefore it
gives only a correction to the chemical potential $\mu$.
The first order anomalous self-energy operator $\hat \Sigma_{20}^{(1)}$
is given by the diagram Fig.2. It is equivalent to the BCS mean field
approximation. Solution of Eq.(\ref{D}) with
$\hat \Sigma_{20}=\hat \Sigma_{20}^{(1)}$  is of the form (see
\cite{Land}):
$\hat G_{\alpha \beta}(p)=\delta_{\alpha \beta}G(p)$,
$\hat F^{\dag}_{\alpha \beta}(p)=g_{\alpha \beta}F^{\dag}(p)$,
$\hat \Sigma_{(20)\alpha \beta}(p)=
g_{\alpha \beta}\Sigma_{20}(p)$, where $\delta_{\alpha \beta}$ and
$g_{\alpha \beta}$ are standard symmetric and antisymmetric spin
matrices, and
\begin{eqnarray}
\label{FGMF}
G(\epsilon,{\bf p})&=&{{u_{\bf p}^2}\over{\epsilon -\epsilon_{\bf p}+i0}}+
{{v_{\bf p}^2}\over{\epsilon +\epsilon_{\bf p}-i0}}\nonumber \\
F^{\dag}(\epsilon,{\bf p})&=&-u_{\bf p} v_{\bf p}
\biggl({1 \over{\epsilon -\epsilon_{\bf p}+i0}}-
{1 \over{\epsilon +\epsilon_{\bf p}-i0}}\biggr) \\
\Sigma^{(1)}_{20}&=&\Delta \nonumber
\end{eqnarray}
with $u_{\bf p},v_{\bf p}=\sqrt{{1\over 2}(1\pm \eta_{\bf p}/
\epsilon_{\bf p})}$. The gap $\Delta$ is given by (\ref{RDS}).

In the next order of perturbation theory
the normal self-energy operator $\Sigma_{11}^{(2)}(p)$ is represented by
the diagrams in Fig.3. Note that that inside $\Sigma^{(2)}$ we use not
the bare Green functions, but
the ``dressed'' ones. They take into account
the self-energy corrections in accordance with Eq.(\ref{D}).
This approach is widely used in  many-body problems
when a correlation correction can be significant, see for
example Ref.\cite{mig}.
The self-energy corresponding to
diagrams in Fig.3 can be easily evaluated.
\begin{equation}
\label{sigma11}
\Sigma^{(2)}_{11}(\epsilon,{\bf p})=\biggl({g\over V}\biggr)^2
\sum_{\bf k_1,k_2}[v_1^2u_2^2-(u_1v_1)(u_2v_2)]
\biggl({{u_3^2}\over{\epsilon -\epsilon_1-\epsilon_2-\epsilon_3}}-
{{v_3^2}\over{-\epsilon -\epsilon_1-\epsilon_2-\epsilon_3}}\biggr)
\end{equation}
The summation here is carried out over ${\bf k_1}$ and ${\bf k_2}$. The
momentum ${\bf k_3}$ is defined as ${\bf k_3=p+k_1+k_2}$. The second order
anomalous self-energy operator is given by the diagrams presented in Fig.4.
The calculation gives
\begin{equation}
\label{sigma20}
\Sigma_{20}^{(2)}(\epsilon,{\bf p})=\biggl({g\over V}\biggr)^2
\sum_{\bf k_1,k_2}u_3v_3 [v_1^2u_2^2-(u_1v_1)(u_2v_2)]
\biggl({1\over{\epsilon -\epsilon_1-\epsilon_2-\epsilon_3}}+
{1\over{-\epsilon -\epsilon_1-\epsilon_2-\epsilon_3}}\biggr)
\end{equation}
Similar to (\ref{sigma11}) the momentum ${\bf k_3}$ in this sum is
equal to ${\bf k_3=p+k_1+k_2}$.

Note that  calculating the first order anomalous self-energy
operator Fig.2 we also have to use the exact Green function.
Hence this operator is proportional to
\begin{equation}
\label{xi}
\Xi^{*}={1 \over V} \int{{d\epsilon}\over {2\pi i}}\sum_{\bf p}
F^{\dag}(\epsilon,{\bf p}),
\end{equation}
where $F^{\dag}$  is the ``dressed'' anomalous Green function, not just
a first order one.

Equations (\ref{D}) can be rewritten as
\begin{eqnarray}
\label{DD}
&&[G^{-1}_0(p)-\Sigma_{11}(p)]G(p)+(g\Xi+\Sigma_{20}^{(2)})
F^{\dag}(p)=1\\
&&[G^{-1}_0(-p)-\Sigma_{11}(-p)]F^{\dag}(p)-
(g\Xi+\Sigma_{20}^{(2)})G(p)=0.\nonumber
\end{eqnarray}
We recall that $\Sigma_{11}^{(1)}$ is momentum independent
and therefore is completely absorbed into the chemical
potential. The second order self-energy (\ref{sigma11}) is
logarithmically divergent at large ${\bf k_2}$. However, the
diverging part is independent of $\epsilon$ and ${\bf p}$, and
therefore can also be absorbed into the chemical potential.
This freedom permits us to renormalize the self-energy imposing the
condition $\Sigma_{11}(\epsilon=0,|{\bf p}|=p_{\mu})=0$ at
$p_{\mu}=\sqrt{2m\mu}$. We can use also the usual linear expansion
near the point $\epsilon=0$, $|{\bf p}|=p_{\mu}$:
$\Sigma_{11}(\epsilon,{\bf p})=
{{\partial \Sigma_{11}}\over{\partial \epsilon}} \epsilon+
{{\partial \Sigma_{11}}\over{\partial \eta_{\bf p}}} \eta_{\bf p}$,
assuming that $\epsilon \sim \eta_{\bf p} \sim \Delta$.
This expansion is certainly valid for the weak coupling
$\Delta \ll E_F$. We have verified numerically using (\ref{sigma11})
that the expansion remains valid with a reasonable accuracy for the strong
coupling $\Delta \sim E_F$ as well. Now we can easily find the solution
of the Eq.(\ref{DD})
\begin{eqnarray}
\label{FGC}
G(\epsilon,{\bf p})&\approx&Z\biggl({{\tilde u_{\bf p}^2}\over{\epsilon
-\tilde \epsilon_{\bf p}+i0}}+ {{\tilde v_{\bf p}^2}\over{\epsilon
+\tilde \epsilon_{\bf p}-i0}}\biggr)\nonumber \\
F^{\dag}(\epsilon,{\bf p})&\approx&-Z {\tilde u_{\bf p}} {\tilde v_{\bf p}}
\biggl({1 \over{\epsilon -\tilde \epsilon_{\bf p}+i0}}-
{1 \over{\epsilon +\tilde \epsilon_{\bf p}-i0}}\biggr) \\
\tilde u_{\bf p},\tilde v_{\bf p}&=&\sqrt{{1\over 2}(1\pm
\tilde \eta_{\bf p}/ \tilde \epsilon_{\bf p})}\nonumber
\end{eqnarray}
where $Z=[1-{{\partial \Sigma_{11}}\over{\partial \epsilon}}]^{-1}$
is the quasiparticle residue, and the renormalized dispersion is
$\tilde \epsilon_{\bf p}= \sqrt{\tilde \eta_{\bf p}^2+\Delta_p^2}$
with
\begin{eqnarray}
\label{tilde}
\Delta_p&=&Z \biggl( g \Xi +\Sigma_{20}^{(2)}(\epsilon,{\bf p}) \biggr) \\
\tilde \eta_{\bf p}&=&Z\biggl(1+{{\partial \Sigma_{11}}\over{\partial
\eta_{\bf p}}}\biggr) \eta_{\bf p} \nonumber
\end{eqnarray}
Numerical computations show that the dependence of $\Delta_p$  on
$\epsilon$ and ${\bf p}$ at
$\epsilon \sim E_F$ and ${\bf p} \sim p_{\mu}$ is actually
rather weak. But anyway, the gap depends on energy and momentum and therefore
it is convenient to introduce $\Delta=\Delta_{\epsilon=0,p=p_{\mu}}$,
and to represent the second order anomalous self-energy as
$\Sigma_{20}^{(2)}(\epsilon,{\bf p})=-\Delta \cdot \sigma(\epsilon,{\bf p})$.
{}From Eqs.(\ref{sigma11}) and (\ref{sigma20}) one finds
\begin{eqnarray}
\label{dsigma}
\biggl({{\partial \Sigma_{11}}\over{\partial \epsilon}}\biggr)_
{\epsilon=0, p=p_{\mu}}&=&-\biggl({{gm}\over{2\pi}}\biggr)^2
R_1(\Delta/\mu),\nonumber\\
\biggl({{\partial \Sigma_{11}}\over{\partial \eta_{\bf p}}}\biggr)_
{\epsilon=0, p=p_{\mu}}&=&\biggl({{gm}\over{2\pi}}\biggr)^2
R_2(\Delta/\mu),\\
\sigma(\epsilon=0, p=p_{\mu})&=&\biggl({{gm}\over{2\pi}}\biggr)^2
R_3(\Delta/\mu).\nonumber
\end{eqnarray}
The functions $R_i,~i=1,2,3$ depend only on the ratio $\Delta/\mu$.
One can easily find that in the weak coupling limit ($\Delta \ll \mu$):
$R_1(0)=const$, $R_2(0)=const$,
$R_3(\Delta / \mu) \approx \ln (\mu /\Delta)$.
Results of numerical
computations of $R_i$ at arbitrary $\Delta / \mu$ are presented in
Table I. They show that the corrections due
to $\Sigma_{11}$ are negligible not only in the weak coupling limit when
$gm/2\pi \ll 1$, $\Delta/\mu \ll 1$, but remain small for the strong
coupling, $gm/2\pi \sim 1$, $\Delta/\mu \ge 1$, as well. Therefore
we can neglect $\Sigma_{11}$  and consider only the anomalous self-energy
operator $\sigma$. The smallness of $\Sigma_{11}$ corrections results in the
fact that $Z \approx 1$. Therefore the exact Green functions (\ref{FGC})
have the form similar to the Green functions (\ref{FGMF}) in the mean-field
approximation.
We used this fact when evaluated Eqs.(\ref{sigma11}),(\ref{sigma20}).

In order to find a relation between the
gap $\Delta$ and the coupling constant $g$ we have to substitute the
solution (\ref{FGC}),(\ref{tilde}) into the self-consistency
condition (\ref{xi}). Then we get
\begin{equation}
\label{eqgap}
1={g\over V}\sum_{\bf p}{1\over{2 \epsilon_{\bf p}}} +
{g\over V}\sum_{\bf p} \int{{d\epsilon}\over{2\pi i}}
{{\sigma(\epsilon,{\bf p})}\over{(\epsilon-\epsilon_{\bf p}+i0)
(\epsilon+\epsilon_{\bf p}-i0)}}.
\end{equation}
The last term here gives the correction to the BCS mean field equation
(\ref{BCS}).
Assuming that the correction to the mean field value of the gap
$\Delta_{mf}$ is small ($\delta \Delta =\Delta -\Delta_{mf} \ll \Delta$)
we find from (\ref{eqgap})
\begin{equation}
\label{deldel}
{{\delta \Delta}\over{\Delta}}=-\biggl({{gm}\over{2\pi}}\biggr)^2
L(\Delta/\mu ),
\end{equation}
where $L$ depends only on the ratio $\Delta/ \mu$
\begin{equation}
\label{L}
L(\Delta/\mu )=-\biggl({{2\pi}\over{gm}}\biggr)^3
{2\over{1+\sqrt{1+\Delta^2/\mu^2}}}
{g\over V}\sum_{\bf p} \int{{d\epsilon}\over{2\pi i}}
{{\sigma(\epsilon,{\bf p})}\over{(\epsilon-\epsilon_{\bf p}+i0)
(\epsilon+\epsilon_{\bf p}-i0)}}.
\end{equation}
Numerical computation of $L$ is straightforward. It is convenient
to integrate in (\ref{L}) along the imaginary $\epsilon$ axis. Results are
presented in Table I. In the weak coupling limit ($\Delta \ll \mu$) $L$
can be easily calculated analytically with logarithmic accuracy:
$L\approx \ln^2(\mu /\Delta)$. Using (\ref{deldel}) and (\ref{RDS})
we find the gap in this limit
\begin{equation}
\label{gap}
\Delta \approx \Delta_{mf}\biggl[1-\biggl({{gm}\over{2\pi}}
\ln {{E_F}\over{\Delta}}\biggr)^2\biggr] \approx
\Delta_{mf}\biggl[1-\biggl({{\ln(E_F/\Delta)}\over{\ln
(2\sqrt{E_FE_{\Lambda}}/\Delta)}}\biggr)^2 \biggr].
\end{equation}
Thus the correction to the mean field value is very large. The
reason for the large correction is simple. The gap is
proportional to $\Delta \propto e^{-2\pi /mg}$. Practically we
have calculated the renormalization of the coupling constant $g$.
A small correction to $g$ gives a large correction to $\Delta$
when exponent $2\pi /mg$ is large. Certainly in this
situation the third order self-energy can give substantial
contribution as well.

 For application to high-$T_c$ superconductivity we are more interested in
the strong coupling limit $\Delta \ge \mu$. Surprisingly in this case the
correction to the mean field approximation is small. For illustration let us
set $E_{\Lambda}/E_F = 20$. With this ratio fixed one can easily find
from (\ref{RDS}) the value of $gm/2\pi $ as a function of $\Delta/\mu$.
After substitution of this value into (\ref{deldel}) with $L$ from
Table I we find $\delta \Delta/\Delta$. Results of these
calculations are presented in Table II.
In our opinion the values of parameters $\Delta/\mu \sim$1--3 and
$E_{\Lambda}/E_F \sim 20$  correspond qualitatively to the $t-J$ model
describing
high-$T_c$ superconductors \cite{Fla4,Bel5}.
We see from Table II that in this region  the correlation correction
$\delta \Delta/\Delta$ is about -10\%.

We conclude that
the BCS-like mean field approximation for the pairing of  dressed holes
in the t-J model
is justified with the accuracy $|\delta \Delta|/\Delta \sim 0.1$.
It is worth to note that the
situation when correlation corrections to the Hartree-Fock
approximation are more important for weak coupling than for
strong coupling is well known for a number of  many-body problems
in nuclear and atomic physics.

\section{The Critical Temperature}

The correction to the critical temperature
may be found in a way similar to the above developed approach
for the correction to the gap at zero
temperature. We have to solve Eqs.(\ref{D}) and find a
point where the gap vanishes. For finite temperature
the energy in these equations is equal
to $\epsilon =i\xi_s$, where $\xi_s=\pi T(2s+1)$, $s=0, \pm 1, \pm 2, ...$
is the Matsubara frequency. Integration over energy inside any
loop should be replaced by summation over Matsubara frequencies.
The mean field approximation is equivalent to the account of the diagram Fig.2
for the self-energy. The solution of
Eqs.(\ref{D}) is of the form:
${\cal G}_{\alpha \beta}(p)=\delta_{\alpha \beta}{\cal G}(p)$,
${\bar {\cal F}}_{\alpha \beta}(p)=-g_{\alpha \beta}
{\bar {\cal F}}(p)$, $\Sigma_{(20)\alpha \beta}(p)=
g_{\alpha \beta}\Sigma_{20}(p)$, where
\begin{eqnarray}
\label{calgf}
{\cal G}(i\xi_s,{\bf p})&=&{1\over{i\xi_s-\eta_{\bf p}}},\nonumber \\
{\bar {\cal F}}(i\xi_s,{\bf p})&=&{{\Delta}\over{\xi_s^2+\eta_{\bf p}^2}},\\
\Sigma_{20}^{(1)}&=&\Delta.\nonumber
\end{eqnarray}
We neglect here all powers of the gap higher than one. Self-consistency
condition for the diagram Fig.2 gives the critical temperature in the mean
field approximation.

 The second order normal self-energy operator is given by the diagrams
presented in Fig.3a,c. The contributions of the diagrams Fig.3b,d,e,f are
proportional to $\Delta^2$ and therefore can be neglected. The calculation
gives
\begin{equation}
\label{sigma11T}
\Sigma^{(2)}_{11}(i\xi_s,{\bf p})=\biggl({g\over V}\biggr)^2
\sum_{\bf k_1,k_2}{{[n(\eta_1)-n(\eta_2)][n(\eta_1-\eta_2)-n(\eta_3]}
\over{i\xi_s +\eta_1-\eta_2-\eta_3}}.
\end{equation}
Here $n(\eta)=[1+\exp (\eta/T)]^{-1}$ is a Fermi-Dirac function.
The summation is carried out over ${\bf k_1}$ and ${\bf k_2}$. The
momentum ${\bf k_3}$ is equal to ${\bf k_3=p+k_1+k_2}$. Similar to the case
of zero temperature the detailed analysis demonstrates that the normal
self-energy operator is small in the strong coupling limit
as well as in the weak coupling one.  Therefore below we
neglect $\Sigma_{11}$.

 The second order anomalous self-energy operator is given by the diagrams
Fig.4a,c,d,e. The contribution Fig.4b is proportional to $\Delta^3$
and therefore is neglected. The calculation gives
$\Sigma_{20}^{(2)}(i\xi,{\bf p})=-\Delta \sigma_T(i\xi,{\bf p})$,
\begin{equation}
\label{sigma20T}
\sigma_T(i\xi_s,{\bf p})=\biggl({g\over V}\biggr)^2
\sum_{\bf k_1,k_2}{{[n(\eta_1)-n(\eta_2)]}\over{2\eta_3}}\biggl(
{{[n(\eta_3)-n(\eta_1-\eta_2)]}\over{i\xi_s +\eta_1-\eta_2-\eta_3}}-
{{[n(-\eta_3)-n(\eta_1-\eta_2)]}\over{i\xi_s +\eta_1-\eta_2+\eta_3}}\biggr),
\end{equation}
where, as above, ${\bf k_3=p+k_1+k_2}$. After substitution
of $\Sigma_{20}^{(2)}$ into (\ref{D}) we find
\begin{equation}
\label{calf}
{\bar {\cal F}}(i\xi_s,{\bf p})={{g\Xi^*-\Delta \sigma_T}\over
{\xi_s^2+\eta_{\bf p}^2}},
\end{equation}
where $\Xi^*=T/V\sum_s \sum_{\bf p}{\bar {\cal F}}(i\xi_s,{\bf p})$ is
given by the diagram Fig.2. Similar to (\ref{eqgap}) this gives the
equation for the critical temperature
\begin{equation}
\label{eqT}
1={g\over V}\sum_{\bf p}{1\over{2 \eta_{\bf p}}}
\tanh {{\eta_{\bf p}}\over {2T_c}} -
{g\over V} T_c \sum_s \sum_{\bf p}{{\sigma_T(i\xi_s,{\bf p})}
\over{\xi_s^2+\eta_{\bf p}^2}}.
\end{equation}
The last term here is the correction to the mean field equation
(\ref{BCS}). Assuming that the correction to the critical temperature
is small we find from (\ref{eqT})
\begin{equation}
\label{delT}
{{\delta T_c}\over{T_c}}=-\biggl({{gm}\over{2\pi}}\biggr)^2
L_T(T_c/\mu_c ).
\end{equation}
The function $L_T$ depends only on the ratio $T_c/ \mu_c$, where $\mu_c$
is the chemical potential at the critical point, and
\begin{equation}
\label{LT}
L_T(T_c/\mu_c)=\biggl({{2\pi}\over{gm}}\biggr)^3
{2\over{1+\tanh (\mu_c/2T_c)}}
{g\over V} T_c \sum_s \sum_{\bf p}{{\sigma_T(i\xi_s,{\bf p})}
\over{\xi_s^2+\eta_{\bf p}^2}}.
\end{equation}
Numerical computation of $L_T$ is straightforward. Results are
presented in the last column of Table I. We present $L_T$ as a
function of $\Delta (0)/\mu (0)$ (the gap at zero temperature over
the chemical potential at zero temperature). It is possible to do so
because $T_c/\mu_c$ itself is a function of $\Delta (0)/\mu (0)$.
In the weak coupling limit $L_T$ can be easily calculated analytically
with logarithmic accuracy: $L_T \approx \ln^2(\mu_c/T_c)\approx
\ln^2(E_F/\Delta (0))$. From (\ref{delT}) and (\ref{RDS}) we find
the critical temperature in this limit
\begin{equation}
\label{TC}
T_c \approx T_{c(mf)}\biggl[1-\biggl({{\ln(E_F/\Delta)}\over
{\ln (2\sqrt{E_FE_{\Lambda}}/\Delta)}}\biggr)^2 \biggr],
\end{equation}
where $T_{c(mf)}$ is the critical temperature in mean field
approximation, and $\Delta =\Delta (0)$ is the gap at zero temperature.
Comparing (\ref{TC}) with (\ref{gap}) we see that $\Delta$ and $T_c$
have the same renormalization factor. Therefore the BCS relation
$\Delta /T_c \approx 1.76$ is preserved despite of the fact that the
renormalizations of the mean field values are about 100\%.

Comparing $L$ and $L_T$ from Table I we see that in the strong
coupling limit $(\Delta \ge \mu)$  the correction to the critical
temperature is larger than the correction to the gap at zero temperature.
Nevertheless the correction remains small. Consider the same example as
for zero temperature: $E_{\Lambda}/E_F = 20$.
The Table II gives $\delta T_c /T_c$ as a function of $\Delta (0)/\mu(0)$.
We see that at $\Delta (0)/\mu(0) \sim$1--3 the correlation correction
$\delta T_c /T_c$ is  about -15\%.

\section{Conclusion}

 We consider the correlation corrections to the BCS mean field
pairing in the two dimensional case. It is found that for the strong pairing
($\Delta \ge \mu$) the correlation corrections are not large: about 10\% of
the corresponding mean field value for the set of parameters relevant
to $t-J$ model describing high-$T_c$ superconductors.
The small values of the corrections is explained qualitatively by the fact
that the energy of virtual excitations becomes higher with increase
of the pairing.
We conclude that the BCS mean field approximation is reasonably justified
for description of the dressed quasiholes pairing in the $t-J$ model.

 Surprisingly for the weak coupling limit ($\Delta \ll \mu$)
the correlation corrections are very large: the renormalization
of the mean field values is about 100\%. The  large
correction results from the  exponential dependence of
the superconducting gap  on
the coupling constant which makes a small correction to $g$ to
give a significant contribution  for  $\Delta$.

\section{ACKNOWLEDGMENTS}

  We are very grateful to M.P.Das, V.V.Flambaum, G.F.Gribakin, and
L.Swierkowski for stimulating discussions.
We acknowledge the Australian National Centre for Theoretical Physics for
organizing the Workshop on High-$T_c$ Superconductivity
at the Australian National University, Canberra 1994,
where this work was began.

\newpage

\tighten

\begin{table}
\caption{Numerical values of dimensionless functions $R_i$, $L$, $L_T$
for different values of $\Delta(T=0)/\mu(T=0)$. We present also the
corresponding
values of $\Delta(T=0)/E_F$. The functions $R_1,R_2$ give normal self-energy
operator $\Sigma_{11}$, see Eq.(14).
Their small values permit one to neglect corrections caused by $\Sigma_{11}$
and consider only  corrections caused by $\Sigma_{20}$. The functions $L$ and
$L_T$ describe the correlation corrections to the gap and to the critical
temperature, see Eqs.(16),(17) and (24),(25).
}
\label{tab1}
\begin{tabular}{cc|cccccccc}
$\Delta/\mu$   & $\Delta/E_F$ & $R_1$  & $R_2$  &  $R_3$   & $L$   &  $L_T$ & &
&  \\
\hline
3 & 1.44 & $3.4\cdot 10^{-2}$ & $1.9\cdot 10^{-2}$ & $0.11$ & $0.27$ & $0.46$
\\
2 & 1.23 & $5.0\cdot 10^{-2}$ & $2.9\cdot 10^{-2}$ & $0.15$ & $0.35$ & $0.59$
\\
1 & 0.83 & $9.8\cdot 10^{-2}$ & $5.8\cdot 10^{-2}$ & $0.31$ & $0.65$ & $0.98$
\\
0.5 & 0.47 & $0.19$ & $0.10$ & $0.66$ & $1.4$ & $1.9$ \\
0.1 & 0.1 & $0.46$ & $0.17$ & $2.1$ & $6.6$ & $7.2$ \\
0.01 & 0.
01 & $0.66$ & $0.24$ & $4.5$ & $23$ & $23$
\end{tabular}
\end{table}

\vspace{20mm}
\begin{table}
\caption{The correlation corrections to the gap at zero temperature and
to the critical temperature for different values of $\Delta(T=0)/\mu(T=0)$.
The ultraviolet cutoff is fixed: $E_{\Lambda}/E_F=20$.}
\label{tab2}
\begin{tabular}{c|cccccccc}
$\Delta/\mu$   & $gm/2\pi$  & $-\delta \Delta /\Delta$ & $-\delta T_c/T_c$ & &
& & & \\
\hline
3 & 0.55 & 0.08 & 0.14 \\
2 & 0.51 & 0.09 & 0.15 \\
1 & 0.42 & 0.12 & 0.18 \\
0.5 & 0.34 & 0.16 & 0.22 \\
0.1 & 0.22 & 0.33 & 0.36 \\
0.01 & 0.15 & 0.50 & 0.50
\end{tabular}
\end{table}

\vspace{20mm}

\begin{center}
{\bf FIGURE CAPTIONS}
\end{center}

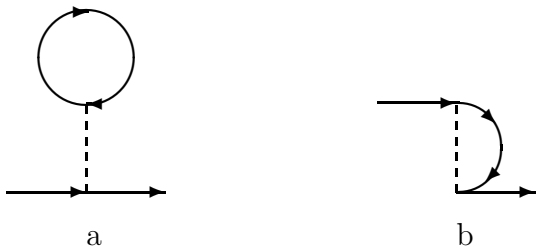
FIG. 1. Normal self energy operator in the first order of perturbation
theory $\Sigma_{11}^{(1)}$.\\

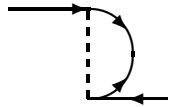
FIG. 2. Anomalous self energy operator in the first order of perturbation
theory $\Sigma_{20}^{(1)}$.\\

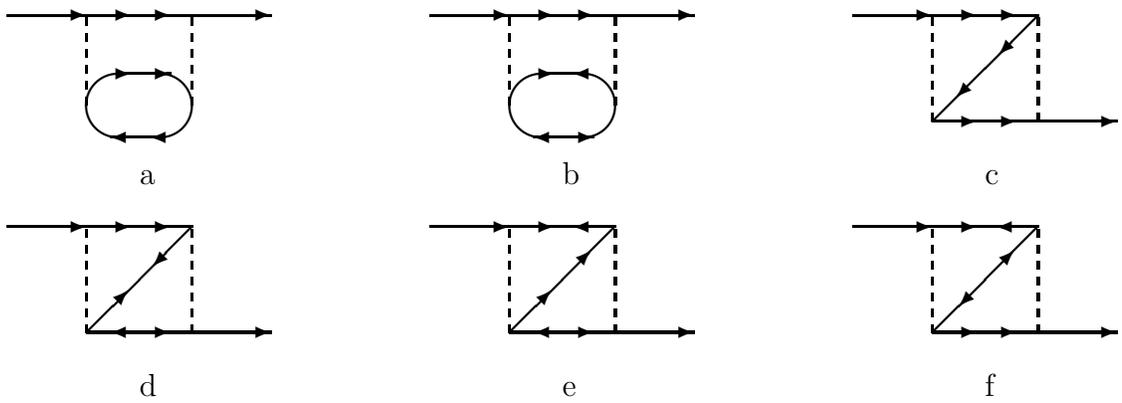
FIG. 3. Normal self energy operator in the second order of perturbation
theory $\Sigma_{11}^{(2)}$.\\

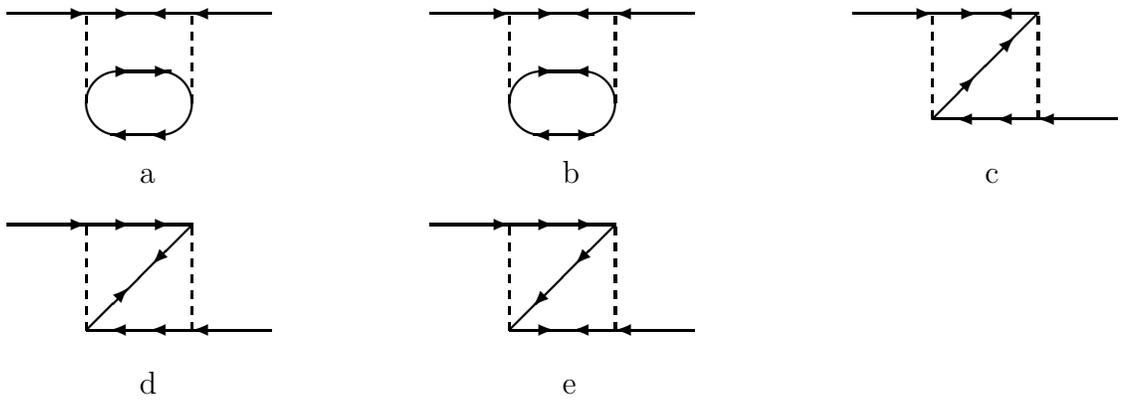
FIG. 4. Anomalous self energy operator in the second order of perturbation
theory $\Sigma_{20}^{(2)}$.

\newpage


\begin{picture}(400,100)
\thicklines
\put(130,40){\vector(1,0){30}}
\put(160,40){\vector(1,0){30}}
\multiput(160,40)(0,6){6}{\line(0,1){3}}
\put(160,91){\circle{34}}
\put(162,73.5){\vector(-1,0){2}}
\put(159,108.5){\vector(1,0){2}}
\put(160,20){a}

\put(270,74){\vector(1,0){30}}
\put(300,40){\vector(1,0){30}}
\multiput(300,40)(0,6){6}{\line(0,1){3}}
\put(300,57){\oval(34,34)[r]}
\put(313,68){\vector(1,-1){2}}
\put(313,46){\vector(-1,-1){2}}
\put(300,20){b}
\put(220,0){Fig.1}
\end{picture}


\begin{picture}(400,100)
\thicklines
\put(190,64){\vector(1,0){30}}
\put(250,30){\vector(-1,0){15}}
\put(235,30){\line(-1,0){15}}
\multiput(220,30)(0,6){6}{\line(0,1){3}}
\put(220,47){\oval(34,34)[r]}
\put(233,58){\vector(1,-1){2}}
\put(233,36){\vector(1,1){2}}
\put(220,0){Fig.2}
\end{picture}


\begin{picture}(440,220)
\thicklines
\put(20,164){\vector(1,0){30}}
\put(50,164){\line(1,0){40}}
\put(65,164){\vector(1,0){2}}
\put(80,164){\vector(1,0){2}}
\put(90,164){\vector(1,0){30}}
\multiput(50,130)(0,6){6}{\line(0,1){3}}
\multiput(90,130)(0,6){6}{\line(0,1){3}}
\put(70,130){\oval(40,24)[r]}
\put(70,130){\oval(40,24)[l]}
\put(65,142){\vector(1,0){2}}
\put(80,142){\vector(1,0){2}}
\put(61,118){\vector(-1,0){2}}
\put(76,118){\vector(-1,0){2}}
\put(70,100){a}

\put(180,164){\vector(1,0){30}}
\put(210,164){\line(1,0){40}}
\put(225,164){\vector(1,0){2}}
\put(240,164){\vector(1,0){2}}
\put(250,164){\vector(1,0){30}}
\multiput(210,130)(0,6){6}{\line(0,1){3}}
\multiput(250,130)(0,6){6}{\line(0,1){3}}
\put(230,130){\oval(40,24)[r]}
\put(230,130){\oval(40,24)[l]}
\put(225,142){\vector(1,0){2}}
\put(240,118){\vector(1,0){2}}
\put(221,118){\vector(-1,0){2}}
\put(236,142){\vector(-1,0){2}}
\put(230,100){b}

\put(340,164){\vector(1,0){30}}
\put(370,164){\line(1,0){40}}
\put(385,164){\vector(1,0){2}}
\put(400,164){\vector(1,0){2}}
\multiput(370,124)(0,6){7}{\line(0,1){3}}
\multiput(410,124)(0,6){7}{\line(0,1){3}}
\put(370,124){\line(1,0){40}}
\put(385,124){\vector(1,0){2}}
\put(400,124){\vector(1,0){2}}
\put(410,124){\vector(1,0){30}}
\put(410,164){\line(-1,-1){40}}
\put(381,135){\vector(-1,-1){2}}
\put(397,151){\vector(-1,-1){2}}
\put(390,100){c}

\put(20,84){\vector(1,0){30}}
\put(50,84){\line(1,0){40}}
\put(65,84){\vector(1,0){2}}
\put(80,84){\vector(1,0){2}}
\multiput(50,44)(0,6){7}{\line(0,1){3}}
\multiput(90,44)(0,6){7}{\line(0,1){3}}
\put(50,44){\line(1,0){40}}
\put(64,58){\vector(1,1){2}}
\put(80,44){\vector(1,0){2}}
\put(90,44){\vector(1,0){30}}
\put(90,84){\line(-1,-1){40}}
\put(61,44){\vector(-1,0){2}}
\put(77,71){\vector(-1,-1){2}}
\put(70,20){d}

\put(180,84){\vector(1,0){30}}
\put(210,84){\line(1,0){40}}
\put(225,84){\vector(1,0){2}}
\put(236,84){\vector(-1,0){2}}
\multiput(210,44)(0,6){7}{\line(0,1){3}}
\multiput(250,44)(0,6){7}{\line(0,1){3}}
\put(210,44){\line(1,0){40}}
\put(250,84){\line(-1,-1){40}}
\put(224,58){\vector(1,1){2}}
\put(239,73){\vector(1,1){2}}
\put(240,44){\vector(1,0){2}}
\put(250,44){\vector(1,0){30}}
\put(221,44){\vector(-1,0){2}}
\put(230,20){e}
\put(220,0){Fig.3}

\put(340,84){\vector(1,0){30}}
\put(370,84){\line(1,0){40}}
\put(385,84){\vector(1,0){2}}
\put(396,84){\vector(-1,0){2}}
\multiput(370,44)(0,6){7}{\line(0,1){3}}
\multiput(410,44)(0,6){7}{\line(0,1){3}}
\put(370,44){\line(1,0){40}}
\put(410,84){\line(-1,-1){40}}
\put(382,56){\vector(-1,-1){2}}
\put(399,73){\vector(1,1){2}}
\put(400,44){\vector(1,0){2}}
\put(385,44){\vector(1,0){2}}
\put(410,44){\vector(1,0){30}}
\put(390,20){f}
\end{picture}


\begin{picture}(440,220)
\thicklines
\put(20,164){\vector(1,0){30}}
\put(50,164){\line(1,0){40}}
\put(65,164){\vector(1,0){2}}
\put(76,164){\vector(-1,0){2}}
\put(120,164){\vector(-1,0){30}}
\multiput(50,130)(0,6){6}{\line(0,1){3}}
\multiput(90,130)(0,6){6}{\line(0,1){3}}
\put(70,130){\oval(40,24)[r]}
\put(70,130){\oval(40,24)[l]}
\put(65,142){\vector(1,0){2}}
\put(80,142){\vector(1,0){2}}
\put(61,118){\vector(-1,0){2}}
\put(76,118){\vector(-1,0){2}}
\put(70,100){a}

\put(180,164){\vector(1,0){30}}
\put(210,164){\line(1,0){40}}
\put(225,164){\vector(1,0){2}}
\put(236,164){\vector(-1,0){2}}
\put(280,164){\vector(-1,0){30}}
\multiput(210,130)(0,6){6}{\line(0,1){3}}
\multiput(250,130)(0,6){6}{\line(0,1){3}}
\put(230,130){\oval(40,24)[r]}
\put(230,130){\oval(40,24)[l]}
\put(225,142){\vector(1,0){2}}
\put(240,118){\vector(1,0){2}}
\put(221,118){\vector(-1,0){2}}
\put(236,142){\vector(-1,0){2}}
\put(230,100){b}

\put(340,164){\vector(1,0){30}}
\put(370,164){\line(1,0){40}}
\put(385,164){\vector(1,0){2}}
\put(396,164){\vector(-1,0){2}}
\multiput(370,124)(0,6){7}{\line(0,1){3}}
\multiput(410,124)(0,6){7}{\line(0,1){3}}
\put(370,124){\line(1,0){40}}
\put(381,124){\vector(-1,0){2}}
\put(396,124){\vector(-1,0){2}}
\put(440,124){\vector(-1,0){30}}
\put(410,164){\line(-1,-1){40}}
\put(384,138){\vector(1,1){2}}
\put(399,153){\vector(1,1){2}}
\put(390,100){c}

\put(20,84){\vector(1,0){30}}
\put(50,84){\line(1,0){40}}
\put(65,84){\vector(1,0){2}}
\put(80,84){\vector(1,0){2}}
\multiput(50,44)(0,6){7}{\line(0,1){3}}
\multiput(90,44)(0,6){7}{\line(0,1){3}}
\put(50,44){\line(1,0){40}}
\put(120,44){\vector(-1,0){30}}
\put(90,84){\line(-1,-1){40}}
\put(64,58){\vector(1,1){2}}
\put(77,71){\vector(-1,-1){2}}
\put(76,44){\vector(-1,0){2}}
\put(61,44){\vector(-1,0){2}}
\put(70,20){d}

\put(180,84){\vector(1,0){30}}
\put(210,84){\line(1,0){40}}
\put(225,84){\vector(1,0){2}}
\put(240,84){\vector(1,0){2}}
\multiput(210,44)(0,6){7}{\line(0,1){3}}
\multiput(250,44)(0,6){7}{\line(0,1){3}}
\put(210,44){\line(1,0){40}}
\put(280,44){\vector(-1,0){30}}
\put(250,84){\line(-1,-1){40}}
\put(221,55){\vector(-1,-1){2}}
\put(237,71){\vector(-1,-1){2}}
\put(236,44){\vector(-1,0){2}}
\put(225,44){\vector(1,0){2}}
\put(230,20){e}
\put(220,0){Fig.4}
\end{picture}

\end{document}